\documentclass[english,twocolumn,superscriptaddress,prl,aps,showpacs]{revtex4}
\usepackage[T1]{fontenc}
\usepackage[latin1]{inputenc}
\usepackage{graphicx}
\usepackage{amssymb}

\makeatletter

\providecommand{\boldsymbol}[1]{\mbox{\boldmath $#1$}}

\makeatletter

\makeatletter

\makeatletter

\makeatletter

\usepackage{bm}

\usepackage{dcolumn}

\usepackage{bm}

\usepackage{endnotes}

\usepackage{amsfonts}

\providecommand{\U}[1]{\protect\rule{.1in}{.1in}}
\makeatletter
\makeatletter
\makeatother
\makeatother

\makeatother

\makeatother

\makeatother

\makeatother

\usepackage{babel}
\makeatother

\begin{document}

\author{Alexey A. Kovalev}

\affiliation{Department of Physics, Texas A\&M University, College Station, TX
77843-4242, USA}

\author{Karel Výborný}

\affiliation{Institute of Physics ASCR, Cukrovarnická 10, 162 53 Praha 6, Czech
Republic }

\author{Jairo Sinova}

\affiliation{Department of Physics, Texas A\&M University, College Station, TX
77843-4242, USA}

\affiliation{Institute of Physics ASCR, Cukrovarnická 10, 162 53 Praha 6, Czech
Republic }

\title{Hybrid skew scattering regime of the anomalous Hall effect in Rashba\\
 systems: unifying Keldysh, Boltzmann, and Kubo formalisms.}

\begin{abstract}
We present the analytical description of the anomalous Hall effect
(AHE) in a 2DEG ferromagnet within the Keldysh formalism. These results unify the three linear response approaches to AHE and close the debate on previous discrepancies. 
We are able to identify a new extrinsic
AHE regime dominated by a hybrid skew scattering mechanism. This new
%hybrid skew scattering 
contribution is inversely proportional to the impurity concentration,
resembling the normal skew scattering, {\em but} independent of
the impurity-strength, resembling the side-jump mechanism. Within
the Kubo formalism this regime is captured by higher order diagrams
which, although weak, can dominate when both subbands are occupied; this
regime can be detected by variable remote doping experiments. 
\end{abstract}

\date{\today{}}


\maketitle
The anomalous Hall effect (AHE) has been a subject of fundamental
research in condensed matter physics for many decades. The anomalous
Hall resistivity $\rho_{xy}$ describes the transverse voltage with
respect to the transport direction and depends on the spontaneous
magnetization $M$ along the $z$ direction. The origin of the AHE
lies in the intrinsic band structure properties \citep{Karplus:sep1954},
and extrinsic spin-asymmetric scattering such as skew-scattering \citep{Smit:1955}
and side-jump scattering \citep{Berger:dec1970}.

Even though the AHE has been studied for a long time \citep{Nozieres:1973},
it still remains a controversial theoretical subject due to the difficulty
to obtain agreement between the different linear response calculations
within equivalent systems %described by Rashba model 
\citep{Dugaev:jun2005,Liu:oct2006,Onoda:sep2006,Inoue:jul2006,Sinitsyn:jan2007,Nunner:dec2007,Borunda:aug2007}.
Recently, some consensus has been reached between the diagrammatic
Kubo formalism \citep{Nunner:dec2007} and the Boltzmann approach
\citep{Sinitsyn:jan2007}. Application of the Keldysh formalism to
the problem is relatively new \citep{Liu:oct2006,Onoda:sep2006} and
connection to the previous theories is required. Liu \textit{et al.}
\citep{Liu:oct2006} employ this approach but fail to reproduce the
diagrammatic results \citep{Nunner:dec2007} because the gradient
expansion of the collision integral is not taken into account %, hence omitting an important contribution to the side-jump component 
\citep{Sinitsyn:jan2007}. Onoda \textit{et al.} \citep{Onoda:sep2006}
use the Keldysh technique formulated in a gauge invariant way; however,
employment of the non-chiral basis representation lacks transparency
and only allows for a numerical solution.

In this Rapid Communication, we derive the kinetic equation that takes into account 
both the effects of the Berry curvature and the extrinsic effects.
We solve the quantum Boltzmann equation analytically in the metallic
(weak scattering) regime, finding the Hall current up to zeroth order
in the impurity strength. Employing the chiral basis allows us to
immediately identify semiclassical contributions \citep{Sinitsyn:jan2007}
such as intrinsic, side-jump and skew-scattering and therefore make
a systematic derivation of the Boltzmann semiclassical approach. We
also make a full connection to the results of the previous works using
the Kubo formalism \citep{Inoue:jul2006,Nunner:dec2007,Sinitsyn:jan2007,Tamara08},
hence bringing to an end the long standing theoretical debate within
the weak scattering regime. In addition, we calculate the important
higher order (hybrid) skew-scattering diagrams. In the limit of high density and mobility,
%when both chiral bands are occupied, 
this hybrid-skew-scattering contribution dominates in the metallic
regime and surprisingly has no dependence on the scattering strength
{\em but} it is inversely proportional to the impurity concentration.

The method presented in the following is general, 
%and can be applied to systems with three dimensions and with more complicated Hamiltonians
%similarly to Onoda \textit{et al} \citep{Onoda:sep2006}. 
however, in order to obtain simple analytical results that connect directly
with other microscopic linear response calculations \citep{Inoue:jul2006,Nunner:dec2007,Sinitsyn:jan2007},
we restrict ourselves to 2D Rashba Hamiltonian with additional exchange
field $h$: \begin{equation}
\hat{H}_{R}=\mathbf{{\vec{\pi}}}^{2}/2m+\alpha\mathbf{{\vec{\pi}}}\cdot\boldsymbol{\hat{\sigma}}\times\mathbf{z}-h\hat{\sigma}_{z}+V(\mathbf{r}),\label{Hamiltonian}\end{equation}
 where $\boldsymbol{\hat{\sigma}}$ are Pauli matrices, ${\vec{\pi}}=\mathbf{k}-e\mathbf{A}$,
$\mathbf{A}(t)=-\mathbf{E}t$ describes the external electric field
and $V(\mathbf{r})$ the impurities. Here and throughout the text
we take $\hbar=c=1$. We employ a simplified model of impurity scattering,
particularly %\begin{equation}
$V(\mathbf{r})=V_{0}\sum_{i}\delta(\mathbf{r}-\mathbf{r}_{i})$, %\label{delta}\end{equation}
where $\mathbf{r}_{i}$ describes the positions of randomly distributed
impurities. We also estimate the spin-orbit coupling component of the disorder potential and show to be important
only in the very high density regime \cite{Engel:oct2005,Crepieux:jul2001}.

We start by writing the Dyson equation \cite{Mahan:1990}: \begin{eqnarray}
\left(\begin{array}{cc}
[\hat{G}_{0}^{R}]^{-1}-\hat{\Sigma}^{R} & -\hat{\Sigma}^{K}\\
0 & [\hat{G}_{0}^{A}]^{-1}-\hat{\Sigma}^{A}\end{array}\right)\otimes\left(\begin{array}{cc}
\hat{G}^{R} & \hat{G}^{K}\\
0 & \hat{G}^{A}\end{array}\right)=\check{1},\label{Dyson}\end{eqnarray}
 where R, A, and K stand for retarded, advanced and Keldysh components
of the Green's functions and self-energies, and the subscript $0$
labels the disorder free system. %$\hat{G}^{R(A)}$ and $\hat{G}_{0}^{R(A)}$ describe the retarded
%(advanced) Green's function for the Hamiltonian Eq. (\ref{Hamiltonian})
%with and without impurities, respectively, $\hat{\Sigma}^{R(A)}$
%and $\hat{\Sigma}^{K}$ describe the retarded (advanced) and Keldysh
%components of the self energy and 
The symbol $\otimes$ denotes a convolution (in position, time and
spin). By considering Eq. (\ref{Dyson}) and its conjugate, %and after subtracting the diagonal part from the nondiagonal, 
we arrive at the Kinetic equation \cite{Mahan:1990}: \begin{equation}
\begin{array}{r}
[\hat{G}_{0}^{R}]^{-1}\otimes\hat{G}^{<}-\hat{G}^{<}\otimes[\hat{G}_{0}^{A}]^{-1}=\hat{\Sigma}^{R}\otimes\hat{G}^{<}-\\
\\\hat{G}^{<}\otimes\hat{\Sigma}^{A}+\hat{\Sigma}^{<}\otimes\hat{G}^{A}-\hat{G}^{R}\otimes\hat{\Sigma}^{<}\end{array},\label{Kinetic}\end{equation}
 where $\hat{G}^{<}/\hat{\Sigma}^{<}\equiv(\hat{G}^{K}/\hat{\Sigma}^{K}+\hat{G}^{A}/\hat{\Sigma}^{A}-\hat{G}^{R}/\hat{\Sigma}^{R})/2$.
%$\hat{\Sigma}^{<}\equiv\hat{\Sigma}^{K}+\hat{\Sigma}^{A}-\hat{\Sigma}^{R}$.
The iterative version of Eq. (\ref{Dyson}) corresponding to the repeated
scattering by impurities is % \begin{eqnarray}
$\check{\Sigma}=\check{\Sigma}_{0}\otimes\left[\check{1}+\check{G}\otimes\check{\Sigma}\right]$,
%\label{Scattering}\end{eqnarray}
 where $\check{\Sigma}_{0}$ is the self-energy from a single scattering
event in Keldysh space. The Keldysh component of this equation, gives the relation:
\begin{eqnarray}
\hat{\Sigma}^{<} & = & \left[1+\hat{G}^{R}\otimes\hat{\Sigma}^{R}\right]\otimes\hat{\Sigma}_{0}^{<}\otimes\left[1+\hat{\Sigma}^{A}\otimes\hat{G}^{A}\right]+\nonumber \\
 &  & \hat{\Sigma}^{R}\otimes\hat{G}^{<}\otimes\hat{\Sigma}^{A},\label{Lesser}\end{eqnarray}
 where for a single scattering event we have $\hat{\Sigma}_{0}^{<}=0$.
Equations (\ref{Kinetic}) and (\ref{Lesser}) form a general closed
set of equations for $\hat{G}^{<}$ and are solved in the following. 

In the presence of slowly varying perturbations, it is useful to perform
the Wigner transformation, \textit{viz}. the center-of-mass coordinates
($X=(R,T)$) and the Fourier transform with respect to the relative
coordinates ($k=(\mathbf{k},\omega)$). In such representation, the
convolution of two operators is approximated as % \begin{equation}
$\hat{A}\otimes\hat{B}\thickapprox\hat{A}\hat{B}+{ \frac{i}{2}}\left(\partial_{X}\hat{A}\partial_{k}\hat{B}-\partial_{k}\hat{A}\partial_{X}\hat{B}\right)$,
%\label{Expansion}\end{equation}
 where we use the four vector notations $\partial_{X}\partial_{k}=\partial_{\mathbf{R}}\partial_{\mathbf{k}}-\partial_{\widetilde{{T}}}\partial_{\omega}$
and $\partial_{\widetilde{{T}}}=\partial_{T}+e\mathbf{E}\partial_{\mathbf{k}}$.
%(see \textit{e.g.} Ref. \citep{Mahan:1990}).
Applying this to the Kinetic Eq. (\ref{Kinetic}) we obtain: \begin{equation}
\begin{array}{c}
\left[\hat{H}_{0},\widehat{G}^{<}\right]+{\displaystyle \frac{i}{2}}\left\{ e\mathbf{E}\boldsymbol{\hat{\upsilon}}_{0},\partial_{\omega}\widehat{G}_{eq}^{<}\right\} +ie\mathbf{E}\partial_{\mathbf{k}}\widehat{G}_{eq}^{<}={\textstyle \hat{\Sigma}^{R}}\widehat{G}^{<}-\\
\\\widehat{G}^{<}{\textstyle \hat{\Sigma}^{A}}+{\textstyle \hat{\Sigma}^{<}}\widehat{G}^{A}-\widehat{G}^{R}{\textstyle \hat{\Sigma}^{<}}+{\displaystyle \frac{i}{2}}\left(\left[{\textstyle \hat{\Sigma}^{R}},\widehat{G}_{eq}^{<}\right]_{p}-\right.\\
\\\left.\left[\widehat{G}_{eq}^{<},{\textstyle \hat{\Sigma}^{A}}\right]_{p}+\left[{\textstyle \hat{\Sigma}_{eq}^{<}},\widehat{G}^{A}\right]_{p}-\left[\widehat{G}^{R},{\textstyle \hat{\Sigma}_{eq}^{<}}\right]_{p}\right)\end{array},\label{SKinetic}\end{equation}
 where $[{\textstyle \hat{A}},\widehat{B}]_{p}\equiv(\partial_{X}{\textstyle \hat{A}}\partial_{k}\widehat{B}-\partial_{k}\widehat{A}\partial_{X}{\textstyle \hat{B}})$,
$\boldsymbol{\hat{\upsilon}}_{0}=\partial_{\mathbf{k}}\hat{H}_{0}$,
$\widehat{G}_{eq}^{<}=n_{F}(\widehat{G}^{A}-\widehat{G}^{R})$ and
${\textstyle \hat{\Sigma}_{eq}^{<}}=n_{F}({\textstyle \hat{\Sigma}^{A}}-{\textstyle \hat{\Sigma}^{R}})$.
In deriving Eq. (\ref{SKinetic}), one retains only the first order
terms in electric field $\mathbf{E}$ and use the fact that our system
is homogeneous and stationary ($\partial_{\mathbf{R}}\widehat{G}^{<}=0$,
$\partial_{T}\widehat{G}^{<}=0$).

To establish the connection with the several mechanisms identified semiclassically
when interpreting the AHE, we transform Eq. (\ref{SKinetic}) into
the chiral basis in which $\hat{H}_{0}$ takes the diagonal form $\hat{S}^{\dagger}\hat{H}_{0}\hat{S}=\widehat{1}k^{2}/2m-\widehat{\sigma}_{z}\lambda,\boldsymbol{\hat{\upsilon}}=\hat{S}^{\dagger}\boldsymbol{\hat{\upsilon}}_{0}\hat{S}$,
and:\[
\begin{array}{c}
\hat{S}=\left(\begin{array}{cc}
\cos\theta/2 & \sin\theta/2\\
ie^{i\varphi}\sin\theta/2 & -ie^{i\varphi}\cos\theta/2\end{array}\right)\end{array},\]
 where $\lambda=\sqrt{(\alpha k)^{2}+h^{2}}$, $\cos(\theta)=h/\lambda$ and $\tan(\phi)=k_{y}/k_{x}$.
We first obtain the intrinsic Hall effect by disregarding the collision
integral in the r.h.s. of Eq. (\ref{SKinetic}). In the chiral basis,
only nondiagonal terms of $\widehat{G}^{<}$ give non-zero contributions to the intrinsic AHE and they are: \begin{eqnarray}
G_{+-}^{c<}=ieE\left(i\upsilon_{y}^{+-}\partial_{\omega}\left[n_{F}(A^{+}+A^{-})\right]/2+\right.\nonumber \\
\left.\left(G_{+}^{R}G_{-}^{R}-G_{+}^{A}G_{-}^{A}\right)\upsilon_{y}^{+-}n_{F}\right)/2\lambda\nonumber \\
G_{-+}^{c<}=-ieE\left(i\upsilon_{y}^{-+}\partial_{\omega}\left[n_{F}(A^{+}+A^{-})\right]/2+\right.\nonumber \\
\left.\left(G_{+}^{R}G_{-}^{R}-G_{+}^{A}G_{-}^{A}\right)\upsilon_{y}^{-+}n_{F}\right)/2\lambda,\label{IntSolution}\end{eqnarray}
 where  $G_{\pm}^{R(A)}=1/(\omega-E_{\pm}+(-)i\Gamma_{\pm}^{(*)})$,
$\Gamma_{\pm}=\Gamma\mp\Gamma_{z}{\displaystyle \frac{h}{\lambda}}$,
$E_{\pm}=k^{2}/2m\pm\lambda$, $A^{\pm}=i(G_{\pm}^{R}-G_{\pm}^{A})$
and $\hat{G}^{c<}$ is $\hat{G}^{<}$ in the chiral basis ($\Gamma$
and $\Gamma_{z}$ are defined below; however, they don't affect the
intrinsic current in the vanishing $\Gamma$ limit). The Green's function $\widehat{G}^{c<}$
allows us to find the intrinsic Hall current along the $x$- axis:\begin{eqnarray}
 & j_{x}=-ie{\displaystyle {\displaystyle \intop}}{\displaystyle \frac{d^{2}\mathbf{k}}{(2\pi)^{2}}}{\displaystyle \frac{d\omega}{2\pi}}{\rm Tr}[\widehat{G}^{c<}\hat{\upsilon}_{x}]=\label{Current}\\
 & -ie^{2}E{\displaystyle \int}{\displaystyle \frac{d^{2}\mathbf{k}}{(2\pi)^{2}}{\displaystyle \frac{d\omega}{2\pi}}n_{F}\frac{(\upsilon_{y}^{+-}\upsilon_{x}^{-+}-\upsilon_{y}^{-+}\upsilon_{x}^{+-})(A^{+}-A^{-})}{4\lambda^{2}}}\nonumber \\
 & =E{\displaystyle \frac{e^{2}}{4\pi}}\left(1-\frac{h}{\lambda_{-}}-(1-\frac{h}{\lambda_{+}})\theta(\omega_{F}-h)\right),\label{Intrinsic}\end{eqnarray}
 where $\lambda_{\pm}=\sqrt{(\alpha k_{\pm})^{2}+h^{2}}$ and $k_{\pm}^{2}=2m (\omega_{F} \mp \lambda_{\pm})$
describe Fermi vectors for the lower/upper chiral bands.

The intrinsic solution Eq. (\ref{IntSolution}) contains both the
contribution at the Fermi level and from the Fermi sea, often referred
to as $\sigma_{xy}^{II}$ conductivity within the Kubo-Streda formalism.
Our next aim is to find the contributions that arise due to impurity
scattering at the Fermi level. We separate Eq. (\ref{SKinetic}) into
two parts, one is proportional to $n_{F}$ and the other is proportional
to $\partial_{\omega}n_{F}$. %The latter part is of interest for the rest of our discussion. 
The part proportional to $\partial_{\omega}n_{F}$, i.e. the Fermi-surface, is: \begin{equation}
\begin{array}{c}
\left[\hat{H}_{0},\widehat{G}^{<}\right]-{\displaystyle \frac{\partial_{\omega}n_{F}}{2}}\left\{ e\mathbf{E}\boldsymbol{\hat{\upsilon}}_{0},\hat{A}\right\} ={\textstyle \hat{\Sigma}^{R}}\widehat{G}^{<}-\widehat{G}^{<}{\textstyle \hat{\Sigma}^{A}}+{\textstyle \hat{\Sigma}^{<}}\widehat{G}^{A}-\\
\\ \widehat{G}^{R}{\textstyle \hat{\Sigma}^{<}}-{\displaystyle \frac{\partial_{\omega}n_{F}}{2}}( \hat{\Gamma}\partial_{\mathbf{k}}\widehat{G}^{A}+\partial_{\mathbf{k}}\widehat{G}^{R}\hat{\Gamma}-\hat{A}\partial_{\mathbf{k}}\widehat{\Sigma}^{A}-\partial_{\mathbf{k}}\widehat{\Sigma}^{R}\hat{A} ) \end{array},\label{SKinetic1}\end{equation}
 where $\hat{A}=i(\hat{G}^{R}-\hat{G}^{A})$ and $\hat{\Gamma}=i({\textstyle \hat{\Sigma}^{R}}-{\textstyle \hat{\Sigma}^{A}})$.
Note that $\partial_{\mathbf{k}}{\textstyle \hat{\Sigma}^{R(A)}}=0$
for the simple delta scatterers. We calculate ${\textstyle \hat{\Sigma}^{R(A)<}}$
and Green's functions $\widehat{G}^{R(A)}$ using the $T$-matrix
approximation up to $n_{i}$-linear terms (in the Pauli basis) \cite{Nunner:dec2007}: %(we find that the second order terms in $n_{i}$ do not contribute to the Hall current): 
\begin{eqnarray}
{\hat{\Sigma}^{R(A)}}=n_{i}V_{0}^{2}\hat{\gamma}^{(*)}(1-V_{0}\hat{\gamma}^{(*)})^{-1}\equiv\mp i(\Gamma^{(*)}\hat{\sigma}_{0}+\Gamma_{z}^{(*)}\hat{\sigma}_{z}),\\
\hat{\Sigma}^{<}=n_{i}V_{0}^{2}\int{\displaystyle \frac{d^{2}k}{(2\pi)^{2}}}(1-V_{0}\hat{\gamma})^{-1}\hat{G}^{<}(1-V_{0}\hat{\gamma}^{*})^{-1},\label{SelfEnergy}\end{eqnarray}
 where $\hat{\gamma}=\int{\displaystyle {d^{2}k}/{(2\pi)^{2}}}{\textstyle \hat{G}^{R}}\equiv\gamma\hat{\sigma}_{0}+\gamma_{z}\hat{\sigma}_{z}$,
${\textstyle \hat{G}^{R}}=(\omega\hat{1}-\hat{H}_{0}-\hat{\Sigma}^{R})^{-1}$,
%{\displaystyle \frac{(\omega-\frac{k^{2}}{2m}+i\Gamma)\hat{\sigma}_{0}+\alpha k_{y}\hat{\sigma}_{x}-\alpha k_{x}\hat{\sigma}_{y}-(h+i\Gamma_{z})\hat{\sigma}_{z}}{(\omega-\frac{k^{2}}{2m}+i\Gamma)-(h+\Gamma_{z})^{2}-\alpha^{2}k^{2}}}$,
with $\gamma=\gamma^{r}+i\gamma^{i}$, $\gamma_{z}=\gamma_{z}^{r}+i\gamma_{z}^{i}$,
and calculated up to the lowest order: \[
\begin{array}{c}
\gamma^{r}={\displaystyle \frac{m}{4\pi}}\ln{\displaystyle \frac{\left|h^{2}-\omega_{F}^{2}\right|}{k_{0}^{4}/4m^{2}}}+{\displaystyle \frac{\alpha^{2}\ln\left|k_{+}^{2}/k_{-}^{2}\right|}{\pi(k_{+}^{2}-k_{-}^{2})/m^{3}}};\\
\gamma^{i}=-{\displaystyle \frac{\nu_{-}+\nu_{+}}{4}};\gamma_{z}^{r}={\displaystyle \frac{h\ln\left|k_{+}^{2}/k_{-}^{2}\right|}{\pi(k_{+}^{2}-k_{-}^{2})/m^{2}}};\gamma_{z}^{i}={\displaystyle \frac{h}{4}(\frac{\nu_{+}}{\lambda_{+}}-\frac{\nu_{-}}{\lambda_{-}})}\end{array},\]
 where $\nu_{\pm}=\frac{m \lambda_{\pm}}{\lambda_{\pm}\pm \alpha^{2}m}$,  $\nu_{+}=0$ when $\omega_{F}<h$, $\frac{\nu_{+}}{\lambda_{+}}=\frac{\nu_{-}}{\lambda_{-}}$
when $\omega_{F}>h$ and $k_{0}$ being the cutoff in the integration
over the $k$ vector. Note that we use the renormalizations $\omega_{F}\rightarrow\omega_{F}-\mbox{Im}\Gamma$
and $h\rightarrow h-\mbox{Im}\Gamma_{z}$ in Eqs. (\ref{SKinetic1},\ref{SelfEnergy})
which allows us to have purely imaginary self energies $\Sigma^{R(A)}$.
However, real parts $\gamma^{r}$ and $\gamma_{z}^{r}$ still appear
in $\hat{\Sigma}^{<}$ in Eq. (\ref{SelfEnergy}). %
\begin{figure}[t]
 %\centerline{\includegraphics[width=1.0\columnwidth]{cartoon.eps}}
\centerline{\includegraphics[scale=0.6]{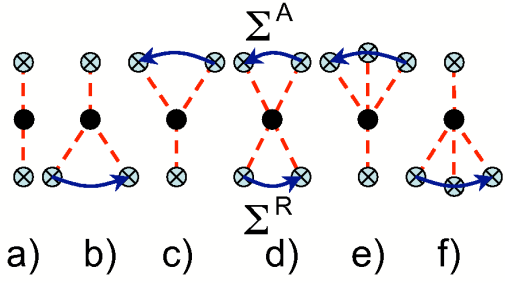}}

\caption{Diagrams representing the averaging procedure in calculating $\hat{\Sigma}^{<}$
in Eq. (\ref{SelfEnergy}) where the upper part of the plot corresponds
to $\hat{\Sigma}^{A}$ and the lower part corresponds to $\hat{\Sigma}^{R}$.
The diagram a) leads to the side-jump contribution and the
disorder-independent skew scattering \cite{SinitsynNote}, the diagrams b) and c) lead to conventional skew
scattering, and the diagrams d), e) and f) lead to the higher order
skew scattering. }

\label{Diagrams} 
\end{figure}

In order to find the current in Eq. (\ref{Current}) up to zeroth
order in $V_{0}$, we transform all elements of Eqs. (\ref{SKinetic1},\ref{SelfEnergy})
into the chiral basis %(including $\hat{T}^{R(A)}$) 
and solve the kinetic equation up to zeroth order in $V_{0}$. That
solution is used to solve the diagonal components of the kinetic equation
up to the second order in $V_{0}$. Note that the expansion of $\hat{G}_{+-/-+}^{c<}$
starts from zero order terms in $V_{0}$ (see Eq. (\ref{IntSolution}))
while the expansion of $\hat{G}_{++/--}^{c<}$ starts from terms proportional
to $V_{0}^{-2}$ which means that we only need to solve the non-diagonal
components of the chiral kinetic equation up to zeroth order in $V_{0}$
while the diagonal components of the kinetic equation has to be solved
up to the second order.

We find different components of $\hat{G}^{c<}=\widehat{G}_{eq}^{c<}+\widehat{G}_{int}^{c<}+\widehat{G}_{(-2)}^{c<}+\widehat{G}_{sj}^{c<}+\widehat{G}_{sk}^{c<}$
in the range (i) $-h<\omega_{F}<h$ when only $"-"$ chiral band is
crossed by the Fermi level. The intrinsic contribution is already
included in Eq. (\ref{Intrinsic}) and its Fermi level part is $\widehat{G}_{int}^{c<}=-iE\partial_{\omega}n_{F}A^{-}(\frac{\alpha\cos\varphi}{4\lambda_{-}}\hat{\sigma}_{x}-\frac{h\alpha\sin\varphi}{4\lambda_{-}^{2}}\hat{\sigma}_{y})$.
By solving the diagonal components of the Kinetic equation up to zeroth
order in $V_{0}$, we obtain the part of $\hat{G}^{c<}$ proportional
to $V_{0}^{-2}$:\[
\widehat{G}_{(-2)}^{c<}=-iE\partial_{\omega}n_{F}A^{-}{\displaystyle \frac{4\lambda_{-}^{2}k_{-}\sin\varphi}{n_{i}V_{0}^{2}\kappa_{-}^{2}\nu_{-}^{2}}}\hat{\sigma}_{--},\]
 where $\kappa_{\pm}=\sqrt{(\alpha k_{\pm})^{2}+4h^{2}}$ and $\hat{\sigma}_{--/++}=(\hat{\sigma}_{0}\pm\hat{\sigma}_{z})/2$.
By solving the non-diagonal components of the Kinetic equation up
to zeroth order in $V_{0}$, we obtain the non-diagonal elements of
the side-jump contribution:\begin{equation}
\begin{array}{c}
\widehat{G}_{sj}^{c<}={\displaystyle \frac{E\partial_{\omega}n_{F}\alpha k_{-}^{2}}{\lambda_{-}}}\left({\displaystyle \frac{(G_{-}^{A}+G_{-}^{R})(\lambda_{-}\hat{\sigma}_{y}\cos\varphi-h\hat{\sigma}_{x}\sin\varphi)}{4\nu_{-}\kappa_{-}^{2}}}+\right.\\
\left.{\displaystyle \frac{iA^{-}(\lambda_{-}\hat{\sigma}_{x}\cos\varphi+3h\hat{\sigma}_{y}\sin\varphi)}{4\nu_{-}\kappa_{-}^{2}}}+{\displaystyle \frac{2iA^{-}h\cos\varphi\hat{\sigma}_{--}}{\kappa_{-}^{2}}}\right)\end{array},\label{sjgap}\end{equation}
 while the diagonal contributions of side-jump are found by considering
the diagram a) in Fig. \ref{Diagrams} and by solving the diagonal
components of the kinetic equation up to the second order in $V_{0}$.
By considering the diagrams b)-f) in Fig. \ref{Diagrams} and by solving
the diagonal components of the kinetic equation up to the second order
in $V_{0}$, we obtain the skew scattering contribution (the last term corresponds to the diagram a) and the disorder-independent skew scattering \cite{SinitsynNote}): \begin{equation}
\begin{array}{c}
\widehat{G}_{sk}^{c<}=iE\partial_{\omega}n_{F}A^{-}\left({\displaystyle \frac{8\Lambda}{n_{i}mV_{0}}}-{\displaystyle \frac{8\gamma^{i}}{n_{i}mV_{0}}}\tan\varphi+\right.\\
\left.{\displaystyle \frac{32\alpha^{2}k_{-}^{2}\gamma^{r}\gamma_{z}^{i}}{n_{i}\kappa_{-}^{2}}}+{\displaystyle \frac{3h\nu_{-}}{\lambda_{-}^{2}}}\right){\displaystyle \frac{\alpha^{2}k_{-}^{3}\lambda_{-}^{2}}{\kappa_{-}^{4}\nu_{-}^{2}}}\hat{\sigma}_{--}\cos\varphi\end{array}\label{skewgap}\end{equation}
 where $\Lambda=(\frac{V_{1}^{3}}{V_{0}^{3}}\gamma_{z}^{i}+\frac{V_{2}^{4}}{V_{0}^{3}}(3\gamma^{r}\gamma_{z}^{i}+\gamma^{i}\gamma_{z}^{r}))$.
Using Eq. (\ref{Current}), we arrive at the the
Hall conductivity \cite{MyNote}:\begin{equation}
\begin{array}{c}
\sigma_{xy}=\sigma_{xy}^{II}+{\displaystyle \frac{e^{2}}{4\pi}}\left({\displaystyle \frac{h\alpha^{2}\nu_{-}}{\lambda_{-}^{2}}}-{\displaystyle \frac{4hk_{-}^{2}\alpha^{2}}{\lambda_{-}\kappa_{-}^{2}}}+\right.\\
\left.{\displaystyle \frac{3hk_{-}^{4}\alpha^{2}}{\kappa_{-}^{4}\nu_{-}}}+{\displaystyle \frac{8k_{-}^{4}\alpha^{2}\lambda_{-}^{2}}{n_{i}V_{0} \kappa_{-}^{4}\nu_{-}^{2}}}({\displaystyle \Lambda}+{\displaystyle \frac{4k_{-}^{2}\alpha^{2}\gamma^{r}\gamma_{z}^{i}}{\kappa_{-}^{2}m}}{\displaystyle \frac{V_{1}^{6}}{V_{0}^{5}}})\right)\end{array},\label{GapHall}\end{equation}
 where $\sigma_{xy}^{II}={ \frac{e^{2}}{4\pi}}(1-\frac{h}{\sqrt{\alpha^{4}+\lambda_{F}^{2}}})$
and $\lambda_{F}=\sqrt{h^{2}+2\alpha^{2}m\omega_{F}}$. In Eqs. (\ref{skewgap},\ref{GapHall})
we have made a straightforward generalization to a more general model
of disorder: $V(\mathbf{r})=\sum_{i}V_{0}^{i}\delta(\mathbf{r}-\mathbf{r}_{i})$
with $\mathbf{r}_{i}$ random and strength distributions satisfying
$\left\langle V_{0}^{i}\right\rangle _{dis}=0$, $\left\langle (V_{0}^{i})^{2}\right\rangle _{dis}=V_{0}^{2}$,
$\left\langle (V_{0}^{i})^{3}\right\rangle _{dis}=V_{1}^{3}$ and
$\left\langle (V_{0}^{i})^{4}\right\rangle _{dis}=V_{2}^{4}$. For
the disorder described after Eq. (\ref{Hamiltonian}), we have $V_{0}=V_{1}=V_{2}$
and for the white noise disorder we have $V_{1}=0$. Note that this
result reduces to the Kubo formalism result of Ref. \citep{Nunner:dec2007}
when the last term bracket is calculated up to zeroth order in the
strength of the disorder. %

We repeat the same procedure in the range (ii) $h<\omega_{F}$ when
both chiral bands are partially occupied. By using Eq. (\ref{Current}),
within this limit we obtain that the intrinsic and side-jump contributions
cancel each other, in agreement with Refs. \onlinecite{Inoue:jul2006,Nunner:dec2007,Sinitsyn:jan2007},
the Fermi sea contribution vanishes ( $\sigma_{xy}^{II}=0$ ) from Eq.
(\ref{IntSolution}), and the Hall conductivity is only non-zero for
the higher order skew scattering arising from diagrams d)-f): \begin{equation}
\begin{array}{l}
\sigma_{xy}=\frac{V_{2}^{4}}{n_{i}V_{0}^{4}}{\displaystyle \frac{e^{2}h\alpha^{2}\ln\left|\frac{k_{-}^{2}}{k_{+}^{2}}\right|}{\pi^{2}(k_{-}^{2}-k_{+}^{2})}}\times\\
\begin{array}{c}
{\displaystyle \frac{\alpha^{2}k_{-}^{2}k_{+}^{2}(k_{+}^{2}-k_{-}^{2})/m+2\sqrt{\alpha^{4}+\lambda_{F}^{2}}(k_{+}^{4}\lambda_{-}-k_{-}^{4}\lambda_{+})}{16(\alpha^{4}+\lambda_{F}^{2})^{3/2}(\lambda_{F}^{2}-h^{2})/\alpha^{2}}}=\end{array} \\  -\frac{V_{2}^{4}}{n_{i}V_{0}^{4}}{\displaystyle \frac{e^{2}h\alpha^{2}\ln\left| k_{-}^{2}/k_{+}^{2} \right|}{\pi^{2}(k_{-}^{2}-k_{+}^{2})}}
\end{array}.\label{MainSkew}\end{equation}
 This contribution from the higher order diagrams d)-f) was not considered
in prior calculations within the Kubo formalism \citep{Inoue:jul2006,Nunner:dec2007}
and only discussed without being calculated in Ref. \onlinecite{Sinitsyn:jan2007}.
We have also used the numerical procedure of Onoda \textit{et al.}
\citep{Onoda:sep2006} to verify this analytical result that identifies
this new extrinsic regime in the 2DEG with Rashba. Although the contribution
in this regime is proportional to $1/n_{i}$ it does not depend on
$V_{0}$ as $V_{0}^{-1}$, as it is usual for the skew scattering.
\begin{figure}[t]
 %\centerline{\includegraphics[width=1.0\columnwidth]{cartoon.eps}}
\centerline{\includegraphics[scale=0.8]{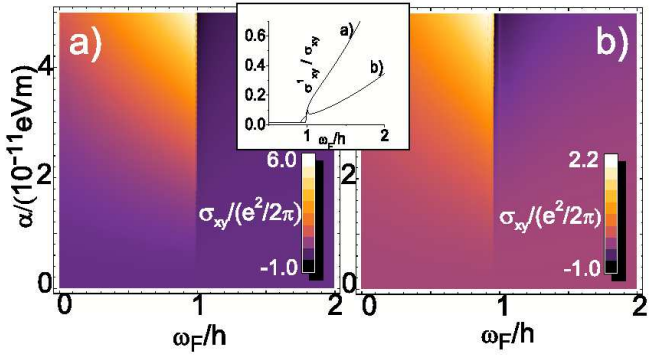}} \vskip -0.3 in
\caption{(Color online.) The anomalous Hall conductivity $\sigma_{yx}$ vs.
the Fermi energy $\omega_{F}$ and the spin-orbit coupling $\alpha$
(concentration of impurities $n_{i}=10^{11}\mbox{cm}^{-2}$, exchange
splitting $h=10\mbox{meV}$ and $m^{*}=0.05m$). The plot a) corresponds
to the mobility $60\:\mbox{m}^{2}/\mbox{Vs}$ or weak impurity scattering
strength while the plot b) corresponds to the mobility $12\:\mbox{m}^{2}/\mbox{Vs}$
and larger impurity strength. In the inset, $\sigma_{xy}^{1}$ corresponds to the extrinsic skew scattering.}
\label{hall} \vskip -0.2 in 
\end{figure}
The result in Eq. (\ref{MainSkew}) can be understood within the Boltzmann
approach by writing the scattering rates in the chiral basis:\begin{equation}
\omega_{\sigma\sigma'}(k,k')=\frac{2 \pi}{\hbar} n_{i}T_{c}^{R}(k,k')_{\sigma\sigma'}T_{c}^{A}(k',k)_{\sigma'\sigma} \delta(\epsilon_{k}-\epsilon_{k'}),\label{Rates}\end{equation}
 where $\hat{T}_{c}^{R(A)}(k,k')=\hat{S}^{\dagger}(k)(1-V_{0}\hat{\gamma}^{(*)})^{-1}\hat{S}(k')$.
The asymmetric part with respect to $k$, $k'$ in Eq. (\ref{Rates})
is responsible for the skew scattering and is proportional to $n_{i}V_{0}^{3}\Lambda$.
Consequently, the Hall current should be proportional to \[
\sigma_{xy}^{{\rm skew}}\thicksim n_{i}V_{0}^{3}\Lambda(\tau_{\pm}^{tr})^{2}\thicksim\frac{n_{i}V_{0}^{3}(\gamma_{z}^{i}+V_{0}(3\gamma^{r}\gamma_{z}^{i}+\gamma^{i}\gamma_{z}^{r}))}{(n_{i}V_{0}^{2})^{2}},\]
 where $\tau_{\pm}^{tr}$ is the transport time for the $\pm$ chiral
bands. The conventional skew scattering ($V_{0}^{-1}$ order) appears
due to the difference in the life-time for the $\pm$ chiral bands
given by $\gamma_{z}^{i}$. However, for the Rashba model when both
subbands are partially occupied we have $\gamma_{z}^{i}=0$. In this
limit, the asymmetry in the scattering still appears due to the difference
in the Fermi energy renormalization for the $\pm$ chiral bands given
by $\gamma_{z}^{r}$ and leads to a $V_{0}$ independent contribution
proportional to $1/n_{i}$.

In Fig. \ref{hall}, we plot the anomalous Hall conductivity as a
function of the Fermi energy $\omega_{F}$ and the spin-orbit coupling
$\alpha$ for attractive impurities ($V_{0}<0$). We take typical parameters corresponding to a high quality
2DEG samples: the carrier concentration is in the range $10^{11}\mbox{cm}^{-2}$,
the maximum spin-orbit coupling is $5\times10^{-11}\mbox{eVm}$ and
the mobilities are $12$ and $60\:\mbox{m}^{2}/\mbox{Vs}$. In the
inset of Fig. \ref{hall} we analyze the importance of the extrinsic skew scattering caused by the impurity induced spin-orbit interaction $H_{SO}=\lambda\left(\boldsymbol{\sigma}\times\boldsymbol{\nabla}V\right)\cdot\mathbf{k}$
($\lambda=0.052\mbox{nm}^{2}$ for GaAs \citep{Engel:oct2005}) that is always present in realistic systems. For the estimate we use the corresponding Hall conductivity \citep{Crepieux:jul2001},
$\sigma_{xy}^{1}=\frac{e^{2} \lambda}{16n_{i}V_{0}}\left(\nu_{-}k_{-}^{4}-\nu_{+}k_{+}^{4}\right)$.
This conductivity becomes important for larger carrier
concentrations and there should be a region of cross-over between the hybrid
skew scattering and the extrinsic skew scattering (some interference between 
the two effects may take place).
In the limit (i) ($\omega_{F}<h$), we observe skew scattering behavior
($\sigma_{xy}\sim1/n_{i}V_{0}$) when the inverse Born scattering
amplitude $\tau=1/n_{i}V_{0}^{2}m\gg1/\epsilon_{F}$ ($\epsilon_{F}$
is the Fermi energy measured from the bottom of the band). For smaller
$\tau$, all curves have asymptotic behavior reaching a sum of side-jump
and intrinsic contributions as it can be seen from Eq. (\ref{GapHall})
which represents the cross-over between the intrinsic-side-jump and
extrinsic anomalous Hall effect. In the transition region to the limit
(ii) ($\omega_{F}>h$), we observe a sudden drop of the Hall conductivity
(see Fig. \ref{hall}) with a sign change. The hybrid skew scattering
should be observable in samples with dopants situated closer to the
2DEG to maximize the impurity strength as it can be seen from the
inset of Fig. \ref{hall}. Onoda \textit{et al.} \citep{Onoda:sep2006}
analyze the region of $\tau\epsilon_{F}\sim1$, finding $\sigma_{xy}\sim\sigma_{xx}^{1.6}$
scaling. This region is beyond applicability of our results which
rely on the weak scattering limit $\tau\epsilon_{F}\gg1$, since our
approximations ignore the corrections to the conductivity $\sim1/\tau\epsilon_{F}$
{and} the gradient expansion in this regime is not fully justified.

Summarizing, we analytically calculate the anomalous Hall current
in a 2DEG ferromagnet with spin-orbit interaction using the Keldysh
formalism. Complete agreement to the Kubo formula approach and to
the Boltzmann equation approach is obtained. By considering the higher
order skew-scattering diagrams, we are able to calculate a Hall current
due to a hybrid skew scattering mechanism which is dominant when both
subbands are partially occupied or when the system has white noise
disorder. This particular Hall current does not depend on the impurity
sign and strength.

\begin{acknowledgments}
We gratefully acknowledge discussions with V. Dugaev, J.
Inoue, T. Jungwirth, A. H. MacDonald, Ar. Abanov, G.E.W. Bauer, N. Sinitsyn and S. Onoda. This work was supported by ONR under grant ONR-N000140610122,
by NSF under grant DMR-0547875 by SWAN-NRI and grants KJB100100802, LC510 and AV0Z10100521. J.S. is a
Cottrell Scholar of the Research Foundation. 
\end{acknowledgments}
\bibliographystyle{apsrev}

\bibliographystyle{apsrev}

\begin{thebibliography}{15}
\bibitem{MyNote}
 \bibinfo{comment}{The retarded (advanced) self-energy is calculated here up to the second order in $V_{0}$}.
\end{thebibliography}

\end{document}